# Full characterization and modelling of graded interfaces in a high lattice-mismatch axial nanowire heterostructure

D.V. Beznasyuk,[1] P. Stepanov,[1] J.L. Rouvière,[2] F. Glas,[3] M. Verheijen,[4,5] J. Claudon,[2] M. Hocevar[1]

[1]*Univ. Grenoble-Alpes, CNRS-Institut Néel, 25 av. des Martyrs, 38000 Grenoble, France*

[2]*Univ. Grenoble-Alpes, CEA-IRIG-PHELIQS, 17 av. des Martyrs, 38000 Grenoble, France*

[3]*Centre for Nanoscience and Nanotechnology, CNRS, Université Paris-Sud, Université Paris-Saclay, 10 bd. Thomas Gobert, 91120 Palaiseau, France*

[4]*Eurofins Materials Science Netherlands, High Tech Campus 11, 5656AE Eindhoven, The Netherlands*

[5]*Department of Applied Physics, Eindhoven University of Technology, 5600 MB Eindhoven, The Netherlands*

Abstract

Controlling the strain level in nanowire heterostructures is critical for obtaining coherent interfaces of high crystalline quality and for the setting of functional properties such as photon emission, carrier mobility or piezoelectricity. In a nanowire axial heterostructure featuring a sharp interface, strain is set by the materials lattice mismatch and the nanowire radius. Here, we show that introducing a graded interface in nanowire heterostructures offers an additional parameter to control strain. For a given interface length and lattice mismatch, we first derive theoretically the maximum nanowire radius below which coherent growth is possible. We validate these findings by growing and characterizing various In(Ga)As/GaAs nanowire heterostructures with graded interfaces. Furthermore, we perform a complete chemical and structural characterization of the interface by combining energy-dispersive X-ray spectroscopy and high resolution transmission electron microscopy. In the case of coherent growth, we directly observe that the mismatch strain relaxes elastically on the side walls of the nanowire around the interface area, while the core of the nanowire remains partially strained. Moreover, our experimental data show good agreement with finite element calculations. This analysis confirms in particular that mechanical strain is largely reduced by interface grading. Overall, our work extends the parameter space for the design of nanowire heterostructures, thus opening new opportunities for nanowire optoelectronics.

Introduction

Semiconductor nanowires offer the unique opportunity to realize coherent axial heterostructures which associate materials having vastly different lattice parameters [1–3] or crystalline structures [4,5]. In addition, the nanowire geometry can be adjusted to finely engineer its photonic and electronic properties [6–10]. Nanowires can hence serve to define optical waveguides [10,11] and cavities [12] or transistor channels [13] and quantum dots [11,14]. Brought together, these appealing features promise a wealth of applications in optoelectronics[6,10]. Prototypes of laser diodes [10,12] and quantum light sources[11], white light emitting diodes [15,16], solar cells [7–9] and high efficiency photodetectors [17,18] were recently developed in nanowire heterostructures.

Today, numerous material combinations have been explored to realize axial nanowire heterostructures. [4,19–21] In all cases, the control of the strain level around the interface is critical, because above a certain threshold, elastic energy is plastically released via the formation of dislocations. [22]. Dislocations act as recombination centers for photons and electrons and degrade the materials properties by reducing the light emission or detection efficiency, and the carrier density. One solution to realize defect-free interfaces is to reduce the nanowire lateral dimensions. Indeed, for a given couple of materials A and B, and thus a given lattice mismatch, there exists a critical nanowire radius below which coherent growth of B on top of A is possible regardless of the height of B. The mismatch strain is then partially and elastically relaxed at the nanowire sidewalls. This critical radius, which separates the domains of elastic and plastic strain relaxation, is well understood in the case of a sharp interface. [23–29]

However, sharp interfaces still present a large residual strain, which may be detrimental for applications. For example, the electron-hole wavefunction overlap decreases in InAs/GaAs quantum dot nanowires, resulting in longer exciton lifetimes and non-radiative recombination [30]. High interface strains also lead to potential barriers for charge carriers, which limits their transport in quantum dot nanowire devices [31]. Moreover, a high interfacial strain can enhance piezoelectric effects which degrades performances of nanowire based solar cells [32]. Finally, in the case of large lattice mismatch, coherent growth is only possible for a very limited range of radii, severely limiting the accessible geometries. As an example, for a lattice mismatch of 7%, the critical radius is as low as 10 nm. Implementing graded interfaces, with a smooth chemical profile, offers a solution to overcome these limitations. Despite a few works [32–34], this strategy remains to be thoroughly explored.

In this Letter, we investigate both theoretically and experimentally graded interfaces in axial nanowire heterostructures. A theoretical model specifies the wire radii compatible with coherent growth for various interface lengths and lattice mismatches. While we specifically consider the representative case of interfaces whose chemical profile is described by an error function, these calculations could be easily extended to other profiles. We successfully confront these predictions to experiments realized with the highly mismatched InAs/GaAs material system. The nanowire heterostructures are grown by molecular beam epitaxy (MBE), using a gold droplet as a catalyst. We perform a complete characterization of the interface: the chemical profile is obtained by energy dispersive X-ray spectroscopy (EDX) analysis, while the structural characterization is conducted through high-resolution transmission electron microscopy (TEM). In the case of coherent growth, the maps of the mismatch strain obtained by geometrical phase analysis (GPA) are in excellent agreement with finite element simulations. This analysis confirms in particular that mechanical strain is largely reduced by interface grading. More generally, interface grading constitutes a novel tuning knob to adjust the physical properties of nanowire heterostructures.

Results

As schematized in Figure 1a, we start by considering an infinitely long nanowire oriented along the z direction, with a circular section of radius $R$. The nanowire features a graded interface between two materials A and B. They have different lattice parameters $a$, which leads to the lattice mismatch $\varepsilon_m = (a_B - a_A)/a_A$. The interface is centered at z = 0, and the fractions $n_A$ and $n_B = 1 - n_A$ of the two species follow a smooth profile. For small-scale compositional gradients (on the order of the nanowire diameter), the interface chemical profile is usually well represented by an error function [35–37]. Specifically, we assume:

$$n_B = \frac{1}{2}\left[1 + \text{erf}\left(\frac{z}{L}\right)\right] \text{ with } \text{erf}\left(\frac{z}{L}\right) = \frac{2}{\sqrt{\pi}}\int_0^{\frac{z}{L}} e^{-u^2} du \qquad (1)$$

where $L$ measures the interface length (Figure 1b). For a given couple of materials, and in the framework of linear elasticity, the amplitude of the strain generated around the interface is controlled by the normalized interface length $\alpha = L/R$. The strain is maximum for an abrupt interface ($\alpha = 0$) and rapidly decreases as $\alpha$ increases. Since plastic relaxation occurs via the formation of dislocations above a certain strain threshold, one thus expects that the critical radius $R_c$ below which coherent growth is possible increases with $\alpha$. These qualitative arguments are confirmed by a quantitative analysis, which combines finite element simulations to compute the elastic energy (assuming mechanical isotropy) with analytical calculations of the formation energy of an edge dislocation (more details in Methods). Figure 1c shows the calculated domains of coherent and plastic growth in the $R - \varepsilon_m$ plane. The solid line which separates the two domains corresponds to the calculated critical radius $R_c(\varepsilon_m, \alpha)$. Strikingly, increasing $\alpha$ leads to a dramatic increase in $R_c$. Compositional graded interfaces thus considerably extend the domain where coherent growth is possible: an interface length over tens of nanometers is sufficient to completely suppress the constraint on the nanowire dimensions. This is in contrast to thin film epitaxy[38] where composition graded buffer layers need to be larger than hundreds of nanometers and to selective area growth of planar nanowires where composition graded interfaces of tens of nanometers are not sufficient to release the mechanical strain and suppress threading dislocations at the surface.[39]

In the following we investigate experimentally axial nanowire heterostructures in the In(Ga)As/GaAs material system in order to compare the theoretical predictions for the evolution of $R_c$ versus $\alpha$ with experimental datasets. Interfaces with $\varepsilon_m$ varying from 0% to 7% can be fabricated in the In(Ga)As/GaAs material system thanks to the possibility of creating ternary alloys. We focus here on high-mismatch heterostructures with $\varepsilon_m \geq 6\%$. We grow our In(Ga)As/GaAs nanowire heterostructures by MBE using the gold assisted vapor-liquid-solid mechanism. The nanowire radius is controlled by the catalyst dimensions, while the interface length can be controlled by adjusting the growth conditions. Indeed, interface grading occurs in particle-seeded nanowire systems and is attributed to the solubility of the growth species in the liquid droplet which constitutes a reservoir [40]. This "reservoir effect" can be tuned or suppressed by carefully adjusting the growth parameters and the droplet dimensions to form either sharp or controlled graded interfaces [26,41,42]. In this work, we have grown several nanowire heterostructures with different $R$, $\varepsilon_m$ and $\alpha$ (growth procedure in Ref[43] and Methods). In all cases, we performed a structural and chemical characterization of the interface which we detail for a first sample labelled $NW_1$.

Figure 2a shows an image of $NW_1$, obtained by scanning transmission electron microscopy using the high-angle annular dark-field imaging mode (HAADF-STEM). The HAADF-STEM image yields a nanowire radius of 10.5 nm, and suggests that the position of the interface stands right after the bottleneck visible in Figure 2a. This is confirmed by the energy dispersive X-ray spectroscopy (EDX) line profile measured along the nanowire axis (Figure 2b). The bottom segment is composed of pure GaAs and the upper one is made of an InGaAs ternary alloy with an average indium composition of 0.8. The corresponding lattice mismatch is $\varepsilon_m =$

6%. Note that EDX radial profiles across both the InGaAs segment and the top of the GaAs segment do not reveal any radial shell. As shown in Figure 2c, the chemical profile of the interface is very well reproduced by an error function. The fit of the data to Equation 1 leads to an interface length $L$ of 5.9 nm (Figure 2c) which corresponds to a reduced interface length $\alpha$ = 0.56. The nanowire radius lies well below the critical value $R_c$ = 56 nm, calculated from the calculated $\alpha$ and $\varepsilon_m$ (Figure 1c).

To investigate the crystalline quality of the nanowire heterostructure, we image different areas of $NW_1$ by high resolution HAADF STEM followed by fast Fourier transform. Both GaAs and InGaAs segments have the wurtzite (WZ) crystal structure except for a small zinc blende (ZB) insertion in the interface region (Supporting Information S.1). Figure 3a and 3b are additional HR STEM images in two different orientations and do not reveal any misfit dislocation in the crystal at the InGaAs/GaAs interface. As predicted by our calculations, the crystalline integrity of our nanowire is preserved and the mismatch strain at the interface is relaxed elastically.

Across the interface, the lattice parameters $a$ and $c$ are modified both by compositional changes and by mechanical strain. To map $a$ and $c$, we employ geometric phase analysis (GPA). Specifically, this technique allows to image the $c$- and $a$-lattice strain[44], i.e the $c$- and $a$-lattice deformations with respect to a reference chosen here as unstrained $c$-GaAs and $a$-GaAs: $\Delta c/c = \frac{c-c_{GaAs}}{c_{GaAs}}$ and $\Delta a/a = \frac{a-a_{GaAs}}{a_{GaAs}}$, respectively. [45] To map the $c$-lattice ($a$-lattice) strain around the interface, we use the high resolution $[2\bar{1}\bar{1}0]$ ($[01\bar{1}0]$) HAADF-STEM image shown in Figure 3a (3b). GPA is then performed to the image by applying a mask around the (0002) (($2\bar{1}\bar{1}0$)) Bragg peak in the Fourier transform visible in the inset of Figure 3a (3b) (see Methods). We choose a medium-size mask of $0.35|\vec{g}|$ (with $\vec{g}$ the reciprocal lattice vector) in order to preserve a balance between a good spatial resolution and a high signal-to-noise ratio.[46,47]

Figure 3c (3d) shows the resulting color-coded map of $\Delta c/c$ ($\Delta a/a$) in the $a-c$ plane. The bottom part of the wire corresponds to unstrained GaAs ($\Delta c/c = \Delta a/a = 0\%$). The top part of the nanowire features a maximum deformation of the $c$- and $a$-planes of 6% with respect to GaAs. This value corresponds to unstrained $In_{0.8}Ga_{0.2}As$, as found by EDX, which indicates full relaxation far from the interface. We observe a transition region around the InGaAs/GaAs interface indicating that the lattice is gradually stretched. Importantly, there is no discontinuity (or defects) in the transition regions for $\Delta c/c$ and $\Delta a/a$, confirming the absence of misfit dislocations at the interface.[27,48] The transition region is thicker in the center than on the nanowire edges, showing that the $a$- and $c$-lattice parameters recover faster their unstrained characteristic value near the nanowire sidewalls than at the nanowire center. It is indeed more difficult to release strain in the core of the nanowire than on the free sidewalls.

To get a complementary insight on strain relaxation, we visualize the arrangement of the crystal planes with a numerical moiré technique[49] (Figure 3e and 3f). We obtain a moiré pattern from the geometric phase images of Figure 3a and 3b using Fourier filtering of the (0002) and ($2\bar{1}\bar{1}0$) Bragg peaks respectively. We observe that the distance between planes is larger in the upper segment than in the bottom segment. Far from the interface, the planes are parallel to each other and are strain-free. Near the InGaAs/GaAs interface, at the sidewalls, the planes bend dramatically. This large deformation is due to elastic relaxation of the mismatch strain at the nanowire free surfaces. Note that plane bending is also evidenced in lattice rotations maps obtained by GPA (not shown).

We now quantitatively compare the experimental GPA data to numerical simulations. We first calculate the mechanical strain tensor $\bar{\bar{\varepsilon}}$ around the nanowire interface using a finite element software (COMSOL Multiphysics) and using all the measured characteristics of $NW_1$. We consider a cylindrical wire of radius $R$ = 10.5 nm. To ensure that finite-length effects are negligible, the lengths of the GaAs and InGaAs sections are both much larger than $R$ (200 nm and 60 nm, respectively). We also include the interface chemical profile as determined from the fit to the EDX measurement (Figure 2b). Finally, we take into account the mechanical anisotropy associated with the wurtzite nanowire crystal (see Methods). The $a$-lattice strain is then deduced using the relation $\Delta a/a = [a_{loc}(\varepsilon_{xx} + 1) - a_{GaAs}]/a_{GaAs}$. Here, $a_{loc}$ is the local unstrained lattice parameter, determined from the measured chemical profile in Figure 2c and using a linear interpolation between GaAs and InAs, $a_{GaAs}$ is the unstrained lattice parameter of GaAs and $\varepsilon_{xx} = \frac{a - a_{loc}}{a_{loc}}$ is the mechanical strain along the x axis. Similarly, we have $\Delta c/c = [c_{loc}(\varepsilon_{zz} + 1) - c_{GaAs}]/c_{GaAs}$. In addition, in order to account for the depth of focus of STEM imaging (around 10 nm), the theoretical data are averaged along the nanowire depth (details in Methods).

Figure 4a and 4b compare the experimental and simulated a- and c-lattice strain along the nanowire axis. Without any free parameter, we obtain for $\Delta c/c$ an excellent agreement between the two profiles (the discrepancy is lower than 0.5%). For $\Delta a/a$, the agreement is good, but the theory predicts a slightly sharper transition than observed in the experimental data. We attribute this to the noise in the experimental GPA data (Figure 3d), which slightly blurs the transition. Both $\Delta c/c$ and $\Delta a/a$ increase gradually from 0% (GaAs reference) to about 6% (In$_{0.8}$Ga$_{0.2}$As segment), which is consistent with the calculated lattice mismatch. Furthermore, we also compute theoretical 2D maps of $\Delta c/c$ and $\Delta a/a$ (see Methods and Supporting Information S.2). They both reproduce the features observed in the experimental maps. Overall, this demonstrates that we have a quantitative understanding of the structural properties of the interface.

We build on this understanding to discuss the distribution of mechanical strain around the InGaAs/GaAs interface. Figures 4c and 4d show the calculated strain components $\varepsilon_{zz}$ and $\varepsilon_{xx}$ along the nanowire axis (z). Both components are zero far from the interface, and feature a significant amplitude over a domain which is 30-40 nm long. Its size significantly exceeds the interface length ($L$ = 5.9 nm), and is in fact roughly set by the nanowire diameter, in agreement with the Saint Venant's principle. $\varepsilon_{zz}$ and $\varepsilon_{xx}$ show a maximum around 0.5%, indicating that the mismatch strain is largely decreased but not fully released. Finally, both $\varepsilon_{zz}$ and $\varepsilon_{xx}$ feature large spatial inhomogeneities. In particular, $\varepsilon_{xx}$ presents several longitudinal oscillations between tensile and compressive deformation. Importantly, these marked strain inhomogeneities will introduce a spatial modulation of the band structure [30,31] which should be taken into account in the design of nanowire devices.

Figure 5 illustrates the dramatic influence of interface grading on the strain fields. We consider an InGaAs/GaAs nanowire with the same dimensions and composition as NW$_1$ and plot the maximal values of $\varepsilon_{xx}$, $\varepsilon_{yy}$ and $\varepsilon_{zz}$ as a function of the interface length $L$. In the case of a sharp interface ($L$ = 0) $\varepsilon_{zz}$ and $\varepsilon_{xx}$ reach 1.7% and 2.6%, respectively. A graded interface with $L$ = 5.9 nm (NW$_1$) is already sufficient to decrease $\varepsilon_{zz}$ by a factor of 3, and $\varepsilon_{xx}$ by a factor close to 6. We note here that interface grading has a stronger influence on the transverse strain components. Of course, increasing $L$ leads to a further decrease of the strain but for the investigated interface lengths, the spatial extension of the strained region is roughly the same (Supporting Information S.3). We next consider the hydrostatic strain $\delta\Omega/\Omega = \varepsilon_{xx} + \varepsilon_{yy} + \varepsilon_{zz}$, which has an important impact on the bandgap and the conduction band offsets.[30] Its maximum value is also plotted in Figure 5: it is reduced from 4.2% down to 0.5% as the interface length increases from 0 to 12 nm. Modest interface grading thus already results in a strong reduction of the strain level.

Next, we consider additional nanowire samples to further support the theoretical predictions of the coherent growth domains. The results are summarized in Figure 6, which confronts the theory to experimental results obtained with two families of samples. The first set of nanowires (NW$_2$ to NW$_4$) features $\varepsilon_m$ of 6%, $R$ around 10 nm and $\alpha$ ranging from 0.25 to 0.56. For all these nanowires, the mismatch strain is always elastically released by the sidewalls, as shown on HRTEM images by the absence of dislocations at the interface and on GPA color-coded maps by the curvature of the *a*- and *c*- planes (Supporting Information S.4). As shown in Figure 6, this nanowire family falls in the predicted coherent domain. Then, we grow a second set of InAs/GaAs nanowire samples with $\alpha$ being 0.48 and 0.67 for a unique $\varepsilon_m$ of 7.2% (NW$_5$ and NW$_6$). Despite a smooth interface, we observe by HRTEM that the nanowires present defects at the InAs/GaAs interface. GPA color-coded maps confirm the presence of misfit dislocations and reveal plane bending (Supporting Information S.5). In those nanowires, the mismatch strain is released both via plastic and elastic relaxation. We finally plot the experimental data in Figure 6: these thick nanowires fall in the plastic relaxation region, confirming here as well the predictions.

We now comment on the spatial distribution of the dislocations in nanowires exhibiting plastic relaxation. In ref [48], the relaxation of the lattice mismatch at a sharp InAs/GaAs interface occurred both plastically and elastically. Authors observe that perfect edge dislocations appear exactly at the interface with a spacing of around 5 nm. In our case the dislocations appear randomly at different heights. This shows a relatively complex strain relaxation that could be caused by a cooperative movement of dislocations [38,50]. This point will be investigated in future work.

Our study is of particular significance when it comes to realize optoelectronic devices using semiconductor heterostructures. Material combinations such as, for example, InP/InSb ($\varepsilon_m$ = 10%) and GaN/InN ($\varepsilon_m$ = 11%) are important for photovoltaic and optoelectronic applications but their structural quality and therefore their physical functions suffer from an extremely high lattice mismatch. As seen in previous works, reducing the diameter is not always possible or sufficient to prevent plastic relaxation [48,51]. Thus, the design of nanowire devices with compositionally graded interfaces has the potential to reduce materials constraint on the device dimensions. Importantly, a compositional grading over few nanometers at nanowire interfaces is sufficient to reduce most of the strain without altering the required physical properties.

Conclusion

In conclusion, we fully characterized high lattice mismatch axial In(Ga)As/GaAs heterostructure nanowires featuring graded interfaces. The heterostructure shows a preserved crystalline quality with a mismatch strain released elastically, via plane bending. Full elastic relaxation occurs at the nanowire sidewalls while the remaining strain is localized in the central area of the nanowire, larger than the interface length. Theoretical predictions confirmed by our experimental data show that the domains for coherent growth can be extended using compositional gradients of few nanometers. Beyond the realization of coherent heterointerfaces, interface grading offers an additional tuning knob to control residual strain in the nanowire, and thus to fine-tune its optoelectronic properties.

Methods

*Calculation of the critical radius:* The critical radius is defined as the maximum radius of a semi-infinite nanowire stem on top of which it is possible to coherently grow a semi-infinite deposit. We perform calculations in the framework of linear elasticity and consider a mechanically isotropic material. The nanowire chemical composition follows the error function profile shown in Eq.1. Assuming Vegard's law, the stress-free lattice mismatch thus varies according to $\varepsilon(z) = \frac{1}{2}\varepsilon_m[1 + \mathrm{erf}(z/L)]$ with $L$ quantifying the interface length and $\varepsilon_m$ the lattice mismatch. For a series of misfits, nanowire radii, and interface lengths, we compare the energies of the system in two states, namely (1) with a purely elastic relaxation of the mechanical strain, and (2) with a single dislocation segment lying perpendicular to the nanowire axis. The critical radius is then defined as the radius above which state (2) has an energy lower than state (1). We first calculated the strain, stress fields and the elastic energy in state (1) using the COMSOL finite elements software. We then calculated the energy in state (2) by using the method of Spencer and Tersoff.[52,53] We explored arbitrary locations of the dislocation segment along $z$ as well as in the plane normal to $z$. Although we did not take the crystalline structure into account, we had to define the magnitude and orientation of the Burgers vectors of the dislocations. We considered Burgers vectors corresponding to pure edge and 60° dislocations of the face centered cubic (or zinc blende) structure, assuming $z$ to coincide with a (111) axis. The 60° dislocations systematically give slightly smaller radii than the edge dislocations.

*Nanowire growth:* The InGaAs/GaAs nanowire extensively studied in this Letter ($NW_1$) originates from a sample grown via the Au-assisted vapor-liquid solid mechanism on GaAs (111)B substrates using a molecular beam epitaxy reactor. We dispersed 20 nm gold colloids on an epi-ready sample and immediately loaded the sample in the reactor. After a deoxidation step at 610°C, the GaAs segment was grown at the same temperature for 25 minutes (and during a cooling step for 5 minutes) with a V/III beam equivalent pressure ratio of 30 and a 2D equivalent Ga growth rate of 0.085 nm/s. The InGaAs growth proceeded at 540 °C for 25 minutes with a V/III beam equivalent pressure ratio of 50 and a 2D equivalent In growth rate of 0.05 nm/s. The presence of a ternary alloy in the upper segment is discussed in Ref[43] and Ref[54]. Finally, the sample was cooled down to 300°C under a constant As flux. The growth parameters of the remaining nanowires discussed in this article can be found in Supporting Information S.6. More information on the growth protocols can be found in Ref[43] and Ref[54].

*TEM and GPA:* We imaged the nanowires by transmission electron microscopy. The nanowires were mechanically removed from the substrate surface and deposited onto carbon grids. High resolution scanning TEM (HRSTEM) images were acquired on a probe-corrected JEOL ARM 200F, operated at 200 kV, equipped with a 100 $mm^2$ Centurio SDD EDs detector. For a probe corrected STEM at 200kV, the depth of focus is 11 nm. We then processed HRSTEM images by means of Geometrical Phase Analysis (GPA) using a home-built software to analyze the strain present in the nanowire. To analyze the $a$-lattice strain, we cannot map $\Delta a/a$ using the same $[2\bar{1}\bar{1}0]$ HAADF-STEM image as for the $c$-lattice because the crystalline phase of our nanowire heterostructure switches systematically from wurtzite to zinc blende at the InGaAs/GaAs interface. As the $[2\bar{1}\bar{1}0]_{WZ}/[011]_{ZB}$ projection of the atomic columns allows one to see the stacking of the different planes ABC for wurtzite and ABAB for zinc blende, Fourier filtering of $(01\bar{1}0)$ planes hides the zinc blende $a$-planes along the nanowire axis (Supporting Information S.1). To solve this issue, we rotate the sample holder to image the nanowire along the $[01\bar{1}0]$ zone axis where the two crystal structures are indistinguishable.

*Mechanical simulations:* We consider a finite InGaAs nanowire having the wurtzite structure and take into accound its mechanical anisotropy. The nanowire is modeled as a cylinder ($R$ = 10.5 nm) with the $z$-axis oriented along the [0001] crystallographic direction and with both $x$-axis and $y$-axis defined along $[\bar{2}110]$ and $[01\bar{1}0]$ directions, respectively. The elastic properties of the nanowire vary along the $z$-axis following the gradual transition in atomic composition observed in the experimental EDX profile from Figure 2c. Based on the fit of the EDX profile, both lattice parameters of the wurtzite phase $a$ and $c$, the density, as well as the stiffness coefficients of the material are recalculated at each point along the $z$-axis. We assume that these parameters vary proportionally to an increasing In content in the InGaAs ternary alloy. For example, for a 50% In content, the values of stiffness coefficients are calculated as an average between WZ GaAs an WZ InAs values. The same stands for other material parameters. The values of the material parameters used in the calculations are listed in Table 1.

| | $a$ (Å) | $c$ (Å) | $C_{11} = C_{22} (GPa)$ | $C_{12}$ (GPa) | $C_{13}$ (GPa) | $C_{33}$ (GPa) | $C_{44}$ (GPa) |
|---|---|---|---|---|---|---|---|
| InAs | 4.269 | 7 | 110.3 | 42.8 | 32.1 | 120.9 | 27.3 |
| GaAs | 3.985 | 6.52 | 147.6 | 46 | 33.4 | 160.2 | 42.2 |

Table 1. Values of lattice constants[55,56] and stiffness coefficients[57] for WZ InAs and WZ GaAs.

In a wurtzite crystal structure, the elastic stiffness coefficients tensor is given as follows,[58]

$$\begin{pmatrix} C_{11} & C_{12} & C_{13} & 0 & 0 & 0 \\ C_{12} & C_{11} & C_{13} & 0 & 0 & 0 \\ C_{13} & C_{13} & C_{33} & 0 & 0 & 0 \\ 0 & 0 & 0 & C_{44} & 0 & 0 \\ 0 & 0 & 0 & 0 & C_{44} & 0 \\ 0 & 0 & 0 & 0 & 0 & \frac{C_{11}-C_{12}}{2} \end{pmatrix}$$

In order to account for a gradually increasing lattice mismatch, we introduce a pseudomorphic strain field

$$\varepsilon_0 = \begin{pmatrix} f_a & 0 & 0 \\ 0 & f_a & 0 \\ 0 & 0 & f_c \end{pmatrix}$$

where $f_a$ and $f_c$ are defined in each point $z$ along the wire axis as follows:

$$f_a = \frac{a_{In_xGa_{1-x}As}(z) - a_{GaAs}}{a_{GaAs}} \quad \text{and} \quad f_c = \frac{c_{In_xGa_{1-x}As}(z) - c_{GaAs}}{c_{GaAs}}$$

The strain is then computed by minimizing the total strain energy using the finite element method. The meshing of the finite wire is implemented using 3D tetrahedron elements. A typical mesh contains in total around $1.5 \times 10^5$ domain elements distributed in a wire, with three degrees of freedom per node. We use a nonuniform and adapted element mesh with small element size (0.18 nm) at the material transition region and larger elements (5 nm) everywhere else. Finally, in order to compare quantitatively the profiles of the *a*- and *c*-lattice strain, we need to take into account (1) the depth of focus of the probe corrected STEM (11 nm) and (2) the lateral sampling of the experimental GPA data. Therefore, the simulated 3D data $\Delta c/c$ and $\Delta a/a$ are averaged along the nanowire depth in a first step. From the resulting 2D maps, the corresponding $\Delta c/c$ and $\Delta a/a$ profiles along the nanowire axis (z) are extracted by averaging laterally each data point over 8 nm.

Conflicts of interest

The authors declare no conflict of interest.

Acknowledgements

Assistance with the MBE growth by Yann Genuist is gratefully acknowledged. Martien den Hertog, Eric Robin and Joël Cibert are acknowledged for fruitful discussions. The authors acknowledge financial support by the LABEX LANEF (ANR-10-LABX-51-01), Université Grenoble Alpes (program AGIR), the Thomas Jefferson Fund of the Embassy of France in the United States and the FACE Foundation, as well as the ANR HYBRID (ANR-17-PIRE-0001) and QDOT (ANR-16-CE09-0010-01).

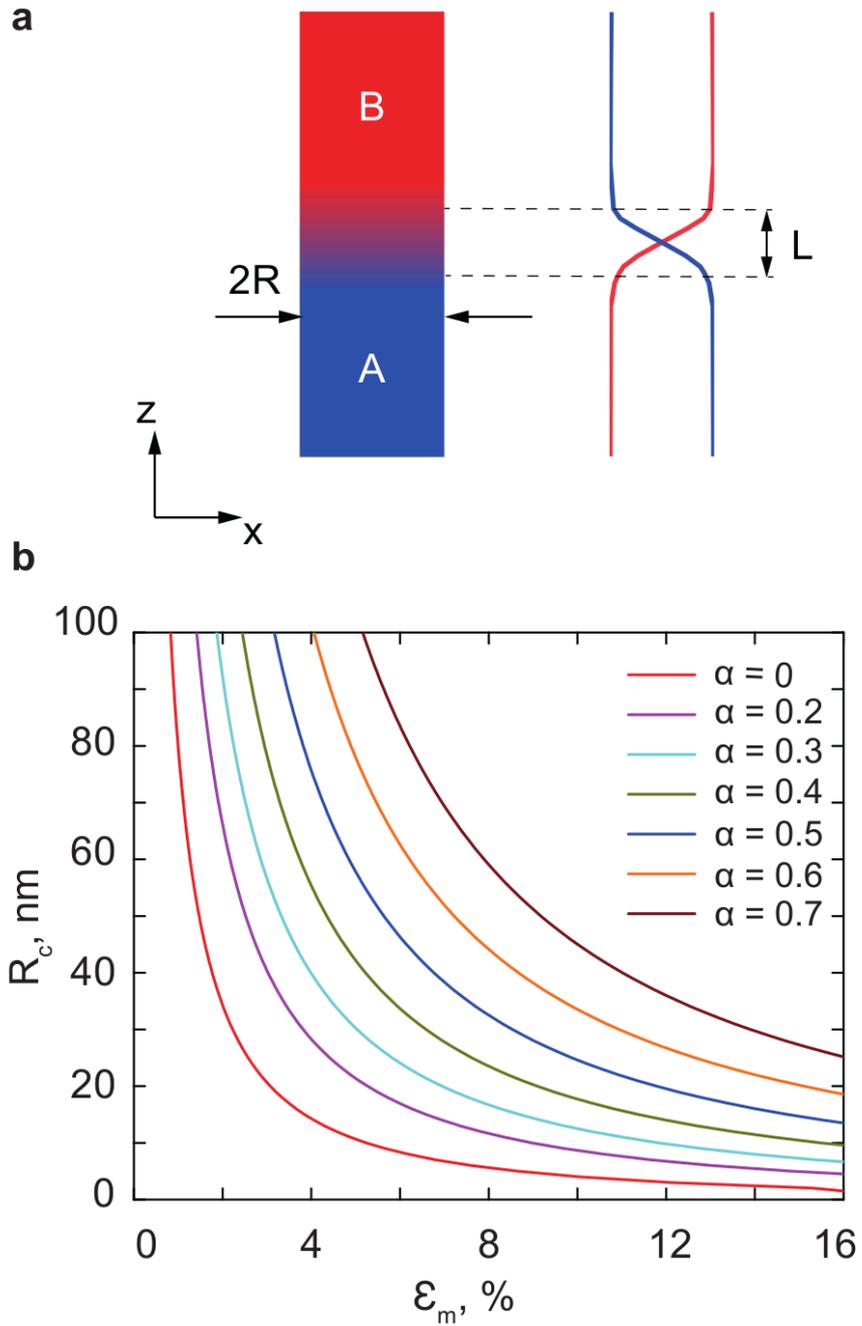

**Figure 1: Axial nanowire heterostructure with graded interfaces. (a)** Schematics of a nanowire heterostructure composed of two materials A and B. The nanowire features a circular section (radius R), its longitudinal axis coincides with the z direction. **(b)** Composition profile along the nanowire axis described by an error function (Eq. 1). Over the interface length $L$, the composition varies by 52% of the total composition jump. **(c)** Calculated critical radius $R_c$ as a function of the mismatch strain $\varepsilon_m$ between A and B for different values of $\alpha = L/R$ ($\alpha = 0$ corresponds to an abrupt interface). For $R < R_c$, growth of B on A is coherent and above $R_c$, plastic relaxation occurs via the introduction of a single edge dislocation segment (see Methods).

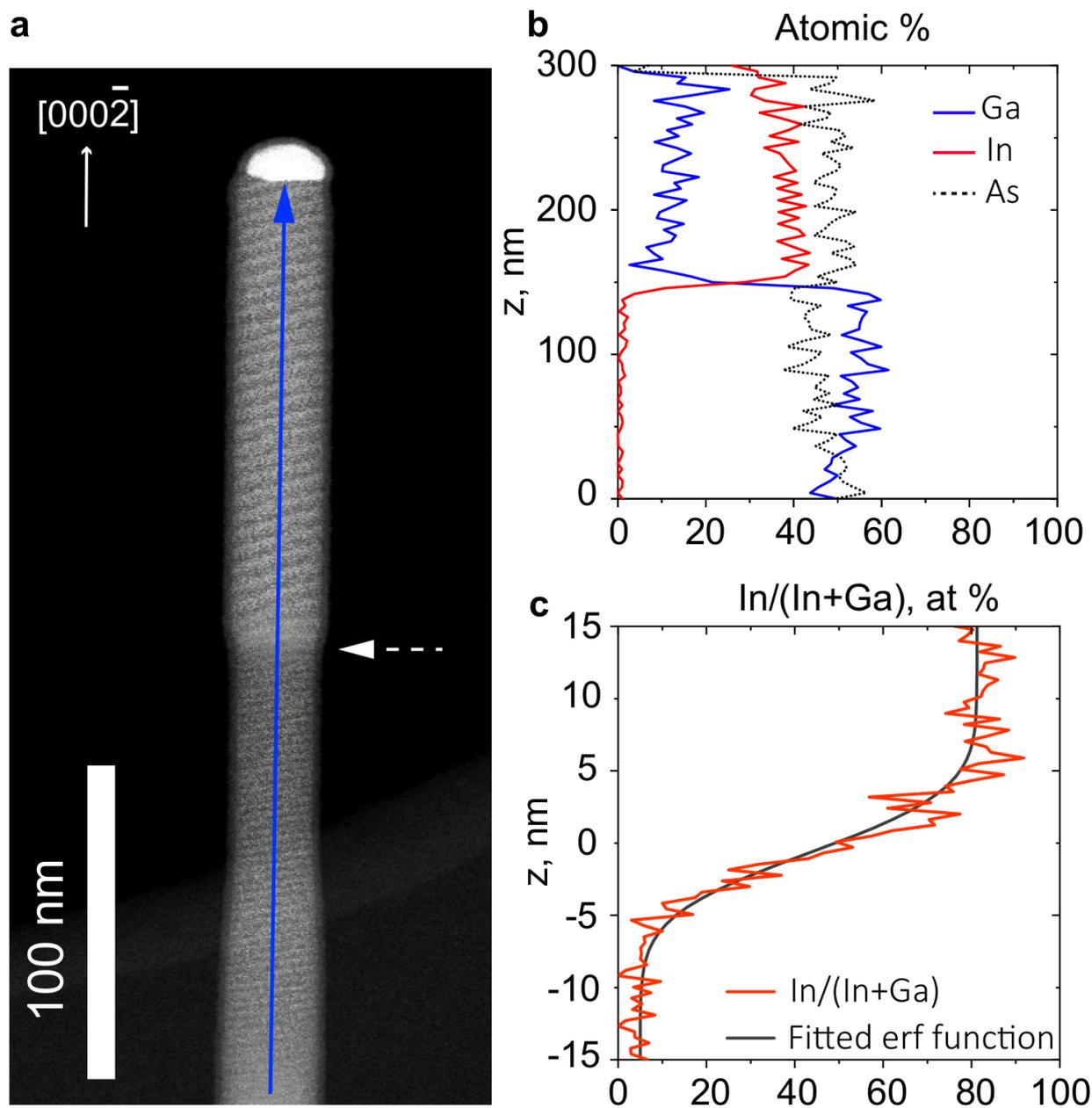

**Figure 2. InGaAs/GaAs axial nanowire heterostructure: chemical characterization. (a)** Dark-field TEM image of NW$_1$ taken along the [2-1-10] zone axis. The position of the interface is indicated by the white arrow. Moiré fringes are visible in the nanowire and are due to the coincidence periods between the scanning step of the electron beam and the interatomic potential. **(b)** EDX composition profile measured along the nanowire axis (blue arrow in (a)). **(c)** Zoom on the interface profile. The fit to an error function yields $L$=5.9 nm.

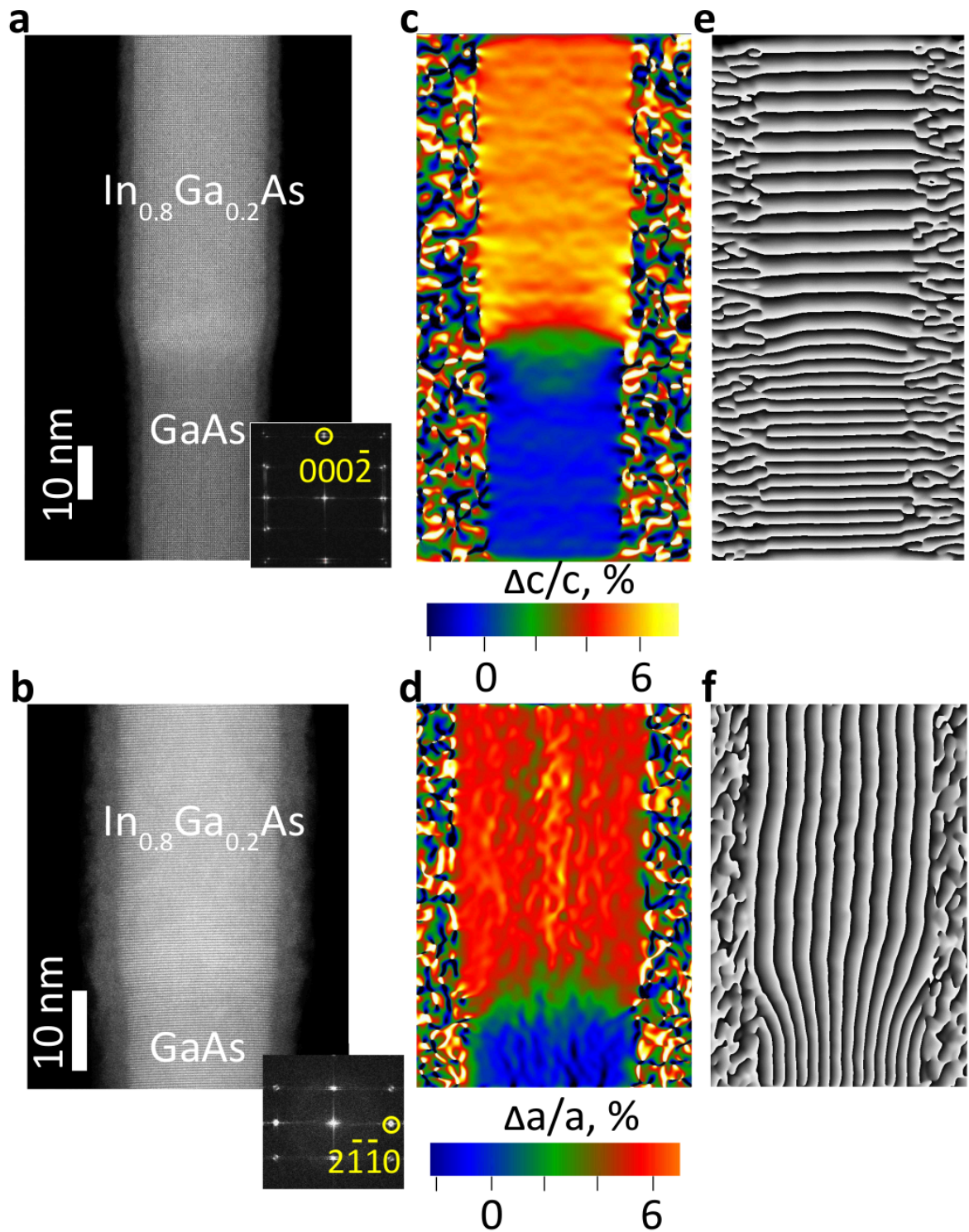

**Figure 3. InGaAs/GaAs axial nanowire heterostructure: high resolution structural characterization.** HAADF STEM image taken along the [2-1-10] viewing direction **(a)** and the [01-10] viewing direction **(b)**. The insets show the corresponding Fast Fourier Transform (FFT). Map of the mismatch strain $\Delta c/c$ **(c)** and $\Delta a/a$ **(d)** obtained by applying GPA on (a) and (b), respectively. Corresponding numerical moiré patterns **(e)** and **(f)**. The scale bars are identical for (a), (c) and (e). Similarly, the scale bars are identical for (b), (d) and (f).

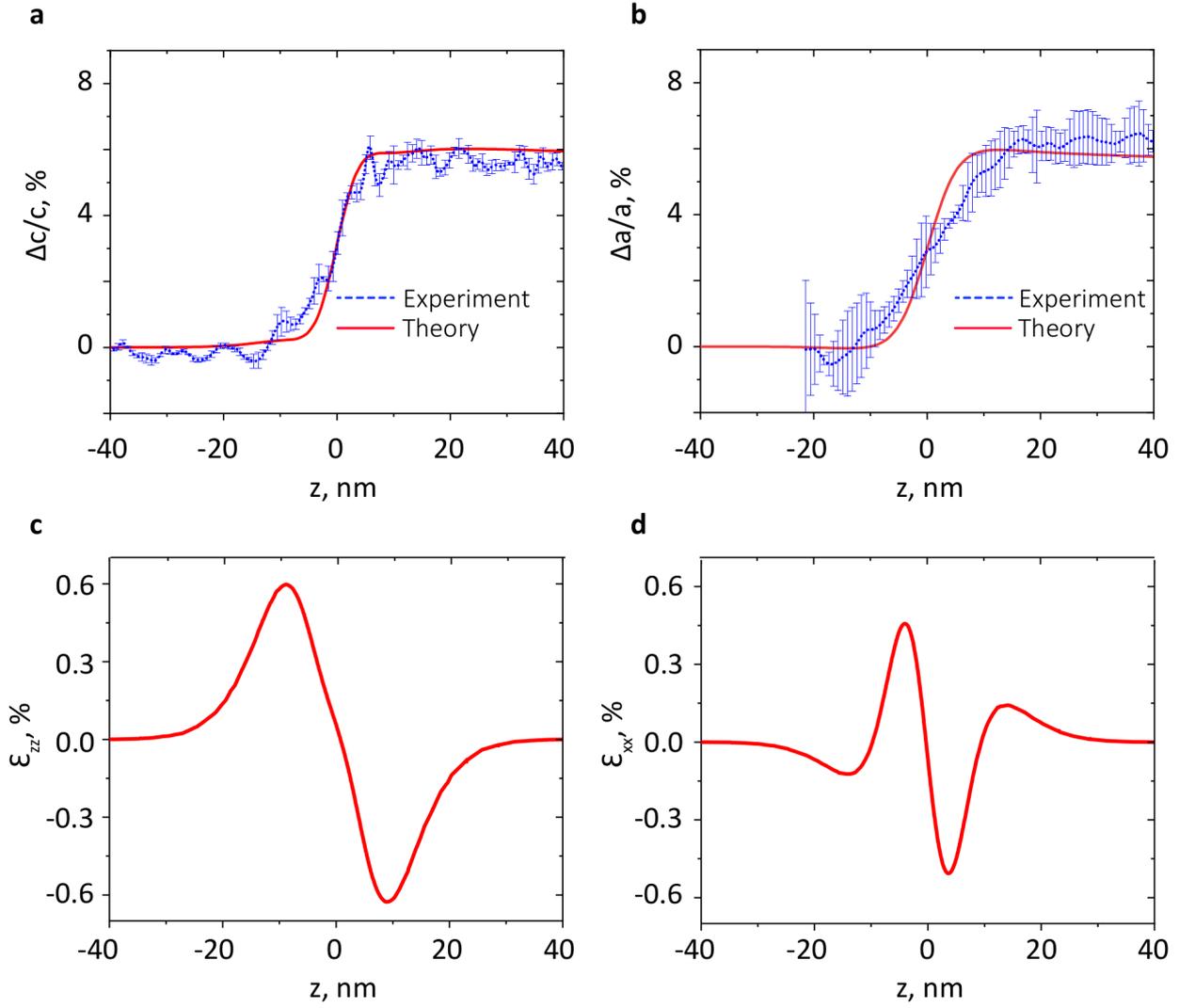

**Figure 4. Experimental and simulated strain profiles along the z axis.** Evolution of $\Delta c/c$ **(a)** and $\Delta a/a$ **(b)** across the interface. The experimental GPA data profiles were extracted from the strain maps and averaged over a lateral sampling of 8 nm. The simulated strains were extracted in the central part of the nanowire and averaged along the electron beam direction. The experimental data appear in blue (with standard deviation as error bars), and the calculation in red. Evolution of the calculated mechanical strain components $\varepsilon_{zz}$ **(c)** and $\varepsilon_{xx}$ **(d)** across the interface.

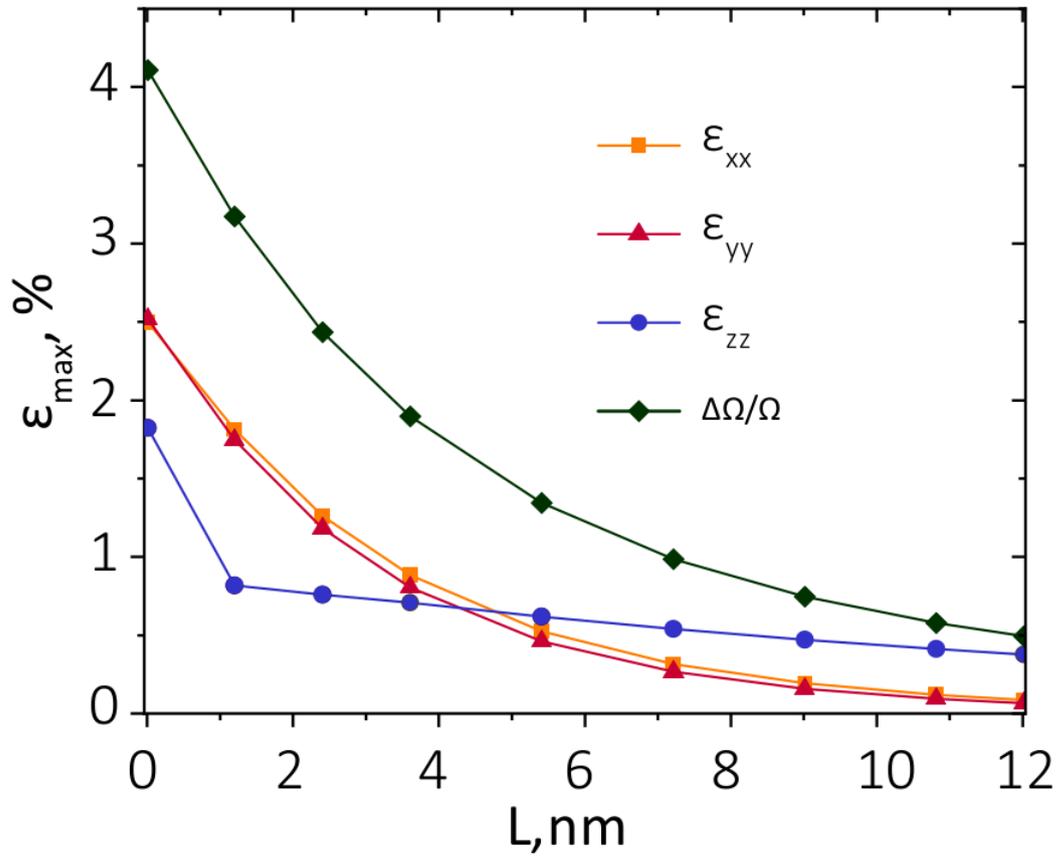

**Figure 5. Influence of the interface length on the mechanical strain**. The maximum hydrostatic strain $\delta\Omega/\Omega$ as well as the maxima of the mechanical strain components $\varepsilon_{xx}, \varepsilon_{yy}$ and $\varepsilon_{zz}$ are plotted versus the interface length $L$.

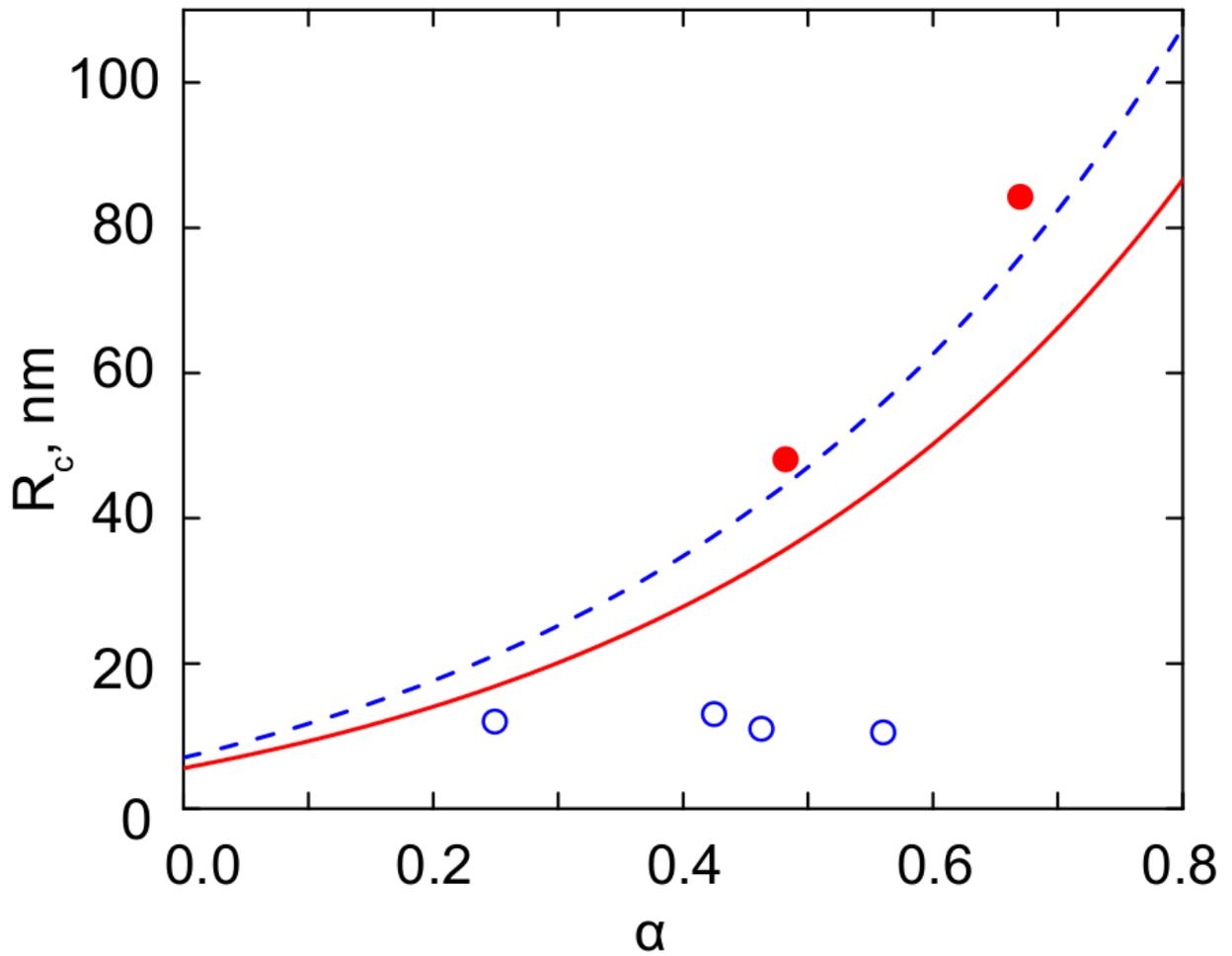

**Figure 6. Elastic and plastic relaxation in axial nanowire heterostructures.** The elastic and plastic domains are separated by a line corresponding to the calculated $R_c$ for $In_{0.8}Ga_{0.2}As/GaAs$ heterostructure (6% lattice mismatch, dashed blue) and InAs/GaAs heterostructure (7.2% lattice mismatch, solid red). Above the line, theory predicts plastic relaxation. The circles indicate the experimental data from dislocation-free (open) and plastically relaxed (solid) structures.